\documentclass[%
 reprint,
 amsmath,amssymb,
pra,
]{revtex4-2}
\usepackage{booktabs}
\usepackage{multirow}
\usepackage{graphicx}
\usepackage{dcolumn}
\usepackage{bm}
\usepackage{subcaption}
\usepackage{ragged2e}  

\begin{document}

\preprint{APS/123-QED}

\title{ Fully passive reference frame independent  quantum key distribution}
\author{Jia-Wei Ying,$^{1}$ Shi-Pu Gu,$^{1}$ Xing-Fu Wang,$^{2}$ Lan Zhou,$^{2}$}\email{zhoul@njupt.edu.cn}
\author{Yu-Bo Sheng$^{1}$}
 \email{shengyb@njupt.edu.cn}
\affiliation{%
 $^1$College of Electronic and Optical Engineering and College of Flexible Electronics (Future Technology), Nanjing
 University of Posts and Telecommunications, Nanjing, 210023, China\\
 $^2$College of Science, Nanjing University of Posts and Telecommunications, Nanjing, 210023, China\\
}%

\date{\today}

\begin{abstract}
Reference-frame-independent quantum key distribution (RFI QKD) significantly alleviates  alignment requirements for reference frame in practical quantum communication systems. While the original protocol requires Alice to prepare six quantum states in   $Z$, $X$, and $Y$ bases, its reliance on active modulation introduces inherent side-channel vulnerabilities from device imperfections. We address this security limitation by integrating a fully passive  source into the RFI framework. In this paper, we propose a fully passive RFI QKD protocol.
Our protocol avoids  active modulation entirely, suppressing side-channel risks through passive quantum state generation.
Moreover, by making full utilization of the quantum states generated by the fully passive source, we enhance the secure key rate of fully passive protocol.
We establish a  system model to  analyze the performance of the protocol. 
Through the optimization of post-selection intervals and intensities, we obtain the maximum secure key transmission rate of the protocol.
Under ideal circumstances, the secure key transmission rate of our protocol can reach more than 50\% of that of the ideal QKD.
Under practical conditions, we have considered the finite-length effect. When  pulse number generated by the  source reaches $10^{12}$,  the maximum communication distances of the protocol can reach 167 km and 136 km with a reference frame misalignment of $0 $ and $45^{\circ} $ respectively.
We believe that our protocol can contribute to the development of practical QKD systems.

\end{abstract}

\maketitle


\section{Introduction}

Quantum key distribution (QKD) enables the sharing of secure keys between two remote users \cite{bb84}. It ensures the unconditional security of communication and has extremely significant applications in quantum communication. In recent years, in order to promote the applications of QKD, many new protocols have been proposed, including: (1). the Measurement Device Independent (MDI) protocol \cite{MDIQKD1,MDIQKD2,PRF,MDIQKD3,MDIQKD4}, which entrusts the measurement to a third party to eliminate the side channels at the measurement end; (2). the  Twin-Field protocol \cite{TWqkd1,TWexp1,TW1,TW2}  and the Mode-Pairing protocol \cite{MPqkd1,MPexp,MP1}  that break the PLOB bound by utilizing  single-photon interference; (3). the Device Independent (DI) protocol \cite{DIQKD1,DIQKD,DIE1,DIE2,DIE3} that eliminates the side channels at the source end and the measurement end by violating Bell's inequality, and so on \cite{CVqkd,qkd7,INI,DRONE}.  

In typical quantum communication scenarios, precise alignment of the reference frames between communicating parties is indispensable for ensuring the accuracy of quantum state measurements. However, practical implementations face inevitable environmental perturbations that induce relative fluctuations in reference frames, making sustained alignment challenging under real-world conditions. To address this limitation,  Laing et al. proposed a Reference Frame Independent (RFI) QKD protocol in 2010 \cite{RFI10}. In their protocol, it is only necessary to ensure the alignment of one axis  of the reference frame. Even if there is a certain misalignment  in the other two axis, a secure key can still be generated. This approach  greatly reduces system complexity. Subsequently, the RFI protocol has been extended  to many QKD protocols \cite{RFIchip,RFIsou,RFIatmo,RFIpm,RFIMDI,RFIMDIfin}.

The original RFI protocol employs    six quantum states prepared  in the $Z_{A}$, $X_{A}$, and $Y_{A}$ bases by Alice. Subsequently, Bob measures these states using the $Z_{B}$, $X_{B}$, and $Y_{B}$ bases \cite{RFI10}. 
Later, some protocols with fewer states were developed \cite{RFI36,RFI44,RFI46,RFI6,RFIMDInn}. Nevertheless, these implementations universally rely on active  source modulation for state preparation – a requirement that introduces critical security vulnerabilities.
The latest research reveals that active modulation operations can generate non-trivial side-channel leakage \cite{sidechannel}, potentially enabling eavesdroppers to deploy sophisticated attacks, including trojan-horse attacks (THA) \cite{trojan0,trojan1,trojan2,trojan3,trojan4,trojan5,trojan6} and unambiguous state discrimination (USD) attacks  \cite{security,USD}, thereby compromising protocol security.

In 2023, Wang et al. proposed a fully passive QKD protocol \cite{FP1} which eliminates active modulation through coherent light interference and combination. 
This architecture passively generates random quantum states and intensities, fundamentally bypassing active-modulation-induced side channels.
The fully passive QKD has been experimentally demonstrated in 2023 \cite{FPexp1,FPexp2} and the method has been applied to a variety of quantum communication protocols \cite{FP3,FP4,FP5,FPCKA,FPMDI1,FPMDI2,FPTW,FPQSDC}.  

Inspired by previous work \cite{RFI10,FP1,FP3,FP4}, we propose a fully passive RFI QKD protocol. 
Our work makes the following key contributions:
(1) We integrate fully passive sources with the RFI protocol, effectively eliminating side-channel vulnerabilities induced by active modulation in original RFI implementations.  By conducting security analysis using the $X$ and $Y$ bases and generating keys with the $Z$ basis, we can fully utilize the quantum states generated  by the fully passive source. This not only enhances practical feasibility for passive QKD systems but also optimizes resource utilization.
(2) We establish a novel system model based on fully passive sources to analyze the performance of the fully passive RFI protocol.   Through the optimization of post-selection intervals and intensities, we obtain the maximum secure key transmission rate that the protocol can achieve and compare it with actively modulated QKD. (3) Taking into account the finite length effect, we analyze the performance of the protocol under practical conditions for different numbers of light source pulses. Furthermore, we also analyzed the performance of the protocol under different  reference frame misalignments. We believe that our work can further promote the practical application of fully passive QKD. 

The structure of the paper is as follows. In Sec. II, we introduce the fully passive RFI QKD, including the protocol, secure key rate, system model, and parameter estimation. In Sec. III, we conduct numerical simulation and parameter optimization. In Sec. IV, we give a  conclusion.

\section{Fully passive RFI QKD }
\subsection{ protocol} \label{sa}
In the protocol, we adopt a fully passive source to prepare the initial state \cite{FP1,FP3,FP4}. Through the interference and combination of coherent light, it is capable of generating coherent pulses with random quantum states and random intensities as $| \sqrt{I}e^{i\psi }  \rangle_{a^{\dagger } (\theta , \phi )}$, where $I$ is the intensity of output pulses and $\psi$ is the global phase. The states in the photon pulses can be expressed as $ a^{\dagger } (\theta , \phi )| vac   \rangle =cos\frac{\theta }{2}| H  \rangle+e^{i\phi}sin\frac{\theta }{2}| V   \rangle$. $\phi$ and $\theta$ are the azimuthal angles on the Bloch sphere. The details are shown in Appendix A.
By measuring the photon pulses in the auxiliary path, we can  filter out the desired output states within the post-selection interval. The post-selection interval is as follows: 
 \begin{eqnarray}
 	S_{Z}^{y}=\{&&\phi \in( 0 , 2 \pi ) ,\\\nonumber
 	&&\theta \in \{ ( \theta _{x} -\bigtriangleup \theta_{Z}   , \theta _{x} +\bigtriangleup \theta_{Z} ) \cap (0,\pi)\},\\\nonumber
 	&&y\in \{vac,d,s\}, \theta _{x} \in \{0, \pi\}
 	\},\\
	S_{X/Y}^{y}=\{&&\phi \in( \phi _{x} -\bigtriangleup \phi _{X/Y} , \phi _{x} +\bigtriangleup \phi _{X/Y} ) ,\\\nonumber
	&&\theta \in ( \frac{\pi}{2}  -\bigtriangleup \theta _{X/Y}   , \frac{\pi}{2} +\bigtriangleup \theta_{X/Y}  ) ,\\\nonumber
	&&y\in \{vac,d,s\}, \phi _{x} \in \{0, \pi\} or \{\frac{\pi}{2}, \frac{3\pi}{2} \}
	\}.
\end{eqnarray}
These three intervals $S_{Z}^{y}$, $S_{X}^{y}$, and $S_{Y}^{y}$ respectively correspond to the $Z$ basis, the $X$ basis, and the $Y$ basis.
When $\theta _{x}=0 (\pi)$, the output state is denoted as  $H$ ($V$). When $\phi _{x}$ takes the values $0,\pi,\frac{\pi}{2}, \frac{3\pi}{2} $, the corresponding  output states are denoted as $D$, $A$, $R$, and  $L$. $y$ represents the intensity intervals $\{vac \ (vacuum), d \ (decoy), s \ (signal)\}$.  $\bigtriangleup \theta_{Z}  $, $\bigtriangleup \theta_{X/Y}  $ and $\bigtriangleup \phi_{X/Y} $ characterize the size of the interval.
The following are the detailed steps of the protocol.

Step 1. Preparing the initial state. Alice first prepares  the initial states with random quantum states and intensities through a fully passive source as $| \sqrt{I}e^{i\psi }  \rangle_{a^{\dagger } (\theta , \phi )}$. 
Through post-selection, Alice filters out the photons within the $S_{Z}^{y}$, $S_{X}^{y}$, and $S_{Y}^{y}$ interval and sends them to Bob. 

Step 2. Measurement. Bob randomly measures the received photons using the  $Z_{B},X_{B},Y_{B}$ bases. Among them, we assume that the $Z_{A(B)}$ basis used by Alice and Bob is well defined, and there is an allowable misalignment $\beta$ between the $X$ and $Y$ bases as:
\begin{eqnarray}
	Z_{B}&&=Z_{A}, \\ \nonumber 
	X_{B}&&=cos\beta X_{A}+sin\beta Y_{A}, \\ \nonumber 
	Y_{B}&&=cos\beta Y_{A}-sin\beta X_{A}.
\end{eqnarray}

Step 3. Parameter estimation. Alice and Bob perform basis comparison. The cases where both of them choose the $Z$ basis are used to generate the key. And the cases where both Alice and Bob choose the $X$ or $Y$ basis are used to estimate Eve's eavesdropping. 
By estimating the security parameters in different cases, Alice and Bob are able to obtain the secure key rate. 

Step 4. Error correction and privacy amplification. Alice and Bob conduct error correction and privacy amplification according to the estimated parameters to obtain the final secure key.

\subsection{secure key rate}
 In Ref. \cite{RFI10},  Laing et al. provide the amount of information that the eavesdropper Eve can intercept in the case of single photons as:
 \begin{eqnarray}
 	IE_{1}=(1-e_{Z_{A}Z_{B}}^{1})h(\frac{1+u}{2} )+e_{Z_{A}Z_{B}}^{1}h(\frac{1+v}{2} ), \label{IE1}
 \end{eqnarray}
where $e_{Z_{A}Z_{B}}^{1}$ is the single-photon error rate when Alice sends photons in the $Z_{A}$ basis and Bob measures them using the $Z_{B}$ basis.  $h(x)$ is defined as the binary Shannon entropy, given by  $h(x)=-x\log_{2}(x)-(1-x)\log_{2}(1-x)$. And 
\begin{eqnarray}
	u&&=min(\frac{\sqrt{C/2}}{1-e_{Z_{A}Z_{B}}^{1} }  ,1),  \label{u}\\
	v&&=\frac{\sqrt{C/2-(1-e_{Z_{A}Z_{B}}^{1})^{2}u^{2}} }{e_{Z_{A}Z_{B}}^{1} }. \label{v}
\end{eqnarray}
Here, $C$ is a parameter that is independent of $\beta$, and is defined as:
\begin{eqnarray}
	C&&=\langle X_{A}X_{B} \rangle^{2} +\langle X_{A}Y_{B} \rangle^{2} +
	\langle Y_{A}X_{B} \rangle^{2} +\langle Y_{A}Y_{B} \rangle^{2} \\ \nonumber
	&&=(1-2e_{X_{A}X_{B}}^{1})^{2}+(1-2e_{X_{A}Y_{B}}^{1})^{2}+\\ \nonumber
	&&\quad (1-2e_{Y_{A}X_{B}}^{1})^{2}+(1-2e_{Y_{A}Y_{B}}^{1})^{2},  \label{C}
\end{eqnarray}
where $e_{\xi_{A}\xi_{B}}^{1}$ is the single-photon error rate when Alice sends photons in the $\xi_{A}$ basis and Bob measures them using the $\xi_{B}$ basis, and $\xi_{A} \in \{ Z_{A},X_{A},Y_{A} \}$, $\xi_{B} \in \{ Z_{B},X_{B},Y_{B} \}$. Since $Z_{A}=Z_{B}$, We abbreviate all of them as $Z$ in the following text.

In an ideal situation, the secure key rate of the fully passive protocol with infinite key can be written as:
\begin{eqnarray}
	R=&& \langle P \rangle_{S_{Z}^{s}} P_{Z} (\langle P(1) \rangle_{S_{Z}^{s}} \langle Y^{1}_{Z} \rangle _{S_{Z}^{s}}
	(1-IE_{1})-\\ \nonumber
	&&\quad\quad\quad\quad\quad f_{e}\langle Q_{Z} \rangle _{S_{Z}^{s}} h(\langle E_{Z} \rangle _{S_{Z}^{s}})),
\end{eqnarray}
where $\langle P \rangle_{S_{Z}^{s}}$ is the probability that Alice generates photons in the interval $S_{Z}^{s}$  and $P_{Z}$ represents the probability that
Bob measures them using the $Z$ basis. 
$\langle P(1)\rangle_{S_{Z}^{s}}$ $ \langle Y^{1}_{Z} \rangle _{S_{Z}^{s}} $ is  the single-photon gain that Alice generates single-photon states in the interval $S_{Z}^{s}$ and Bob selects the Z basis for measurement and obtains a response. $f_{e}$  is the error correction coefficient. $\langle Q_{Z} \rangle _{S_{Z}^{s}}$ ($\langle E_{Z} \rangle _{S_{Z}^{s}}$) is the overall gain (error rate) that Alice generates  states in the interval $S_{Z}^{s}$ and Bob selects the $Z$ basis for measurement and obtains a response.

However, in practical experiments, it is not realistic to generate an infinitely long key. It is necessary to analyze the secure key rate of the system under the condition of a finite key length. According to \cite{FP4}, we present the formula for the secure key length with finite-key analysis as:
\begin{eqnarray}
	l=\lfloor  M_{S_{Z}^{s}Z,1}^{L}(1-IE_{1})-\lambda _{EC}-log(\frac{1}{2\epsilon_{cor}\epsilon_{PA}^{2}\delta  })\rfloor , \label{l}
\end{eqnarray}
where $ M_{S_{Z}^{s}Z,1}^{L}$ is the lower bound of the single-photon response count when generating the key. $IE_{1}$ is the information entropy that Eve can obtain from a single-photon state within the framework of the RFI. $\lambda _{EC}$ represents the information entropy that needs to be consumed for error correction and $\lambda _{EC}=f_{e}M_{S_{Z}^{s}Z}^{U}h(E_{S_{Z}^{s}Z}^{U})$.  Among them, $f_{e}$ is the error correction coefficient, and $M_{S_{Z}^{s}Z}^{U}$ ($E_{S_{Z}^{s}Z}^{U}$) is the upper bound of the total response count (error rate) of the key ($ZZ$) basis.   $\epsilon_{cor}$ ($\epsilon_{PA}$) denotes the probability that error correction (privacy amplification) fails. $\delta $ represents the accuracy of smooth min-entropy estimation.
Since the key length should be an integer, the entire expression ultimately needs to be rounded down.

\subsection{The overall gain $Q$ and error rate $E$}
Here, through system modeling, we  analyze the overall gain $Q$ and error rate $E$ of the protocol under theoretical circumstances. 
According to Appendix A, the pulse emitted by the light source can be expressed as: 
\begin{eqnarray}
	| \mu_{out}   \rangle&&=| \sqrt{I_{e}}e^{i\psi }  \rangle_{a^{\dagger } (\theta , \phi )}=\sum_{n=0}^{\infty }\sqrt{P(I,n)}e^{i\psi n } | n \rangle _{a^{\dagger } (\theta , \phi )}.\quad\quad
\end{eqnarray}
$P(I,n)$  is the probability that the source emits n photons, which follows a Poisson distribution,  and $P(I,n) =e^{-I }\frac{I ^{n}}{n!}$. 
Consider a fiber loss channel with a transmission efficiency of $\eta_{c}=10^{-\frac{\alpha L}{10} }$, where $\alpha$ is the fiber attenuation coefficient and $L$ is the transmission distance. After the output state passes through the channel, it becomes:
 \begin{eqnarray}
 	| \mu  \rangle&&=\sum_{n=0}^{\infty }\sqrt{P(I \eta_{c},n)} e^{i\psi n } | n \rangle _{a^{\dagger } (\theta , \phi )}.
 \end{eqnarray}
The formula can be understood as that at Bob's end, there is a probability of $P(I \eta_{c},n)$ to receive an $n$-photon state $| n \rangle _{a^{\dagger } (\theta , \phi )}$. 
Next, the $n$-photon state will be measured by projection in the $Z_{B},X_{B},Y_{B}$ bases. Note that the eigenstates of the $Z_{B},X_{B},Y_{B}$ bases are $\{ | H \rangle , | V \rangle  \}$, $\{ | D' \rangle = \frac{1}{\sqrt{2} } (| H \rangle +e^{i \beta } | V \rangle), | A' \rangle= \frac{1}{\sqrt{2} } (| H \rangle -e^{i \beta } | V \rangle)  \}$,  $\{ | R' \rangle = \frac{1}{\sqrt{2} } (| H \rangle +i e^{i \beta } | V \rangle), | L' \rangle= \frac{1}{\sqrt{2} } (| H \rangle -i e^{i \beta } | V \rangle)  \}$, respectively.
The $n$-photon state, when expanded in these bases, becomes:
\begin{eqnarray}
	| n \rangle _{a^{\dagger } (\theta , \phi )}=&&\frac{1}{\sqrt{n!} }
	(a^{\dagger } (\theta , \phi ))^{n} | vac \rangle \label{n} \\ \nonumber
	=&&\frac{1}{\sqrt{n!} }(cos\frac{\theta}{2} a_{H}^{\dagger }+
	e^{i\phi  }sin\frac{\theta}{2}a_{V}^{\dagger })^{n}| vac \rangle\\ \nonumber
	=&&\frac{1}{\sqrt{n!} }(f(x) a_{x}^{\dagger }+
	f(y)a_{y}^{\dagger })^{n}| vac \rangle\\ \nonumber
	=&& \frac{1}{\sqrt{n!} }\sum_{m=0}^{n} C_{n}^{m}(f(x)a_{x}^{\dagger })^{m}(f(y)a_{y}^{\dagger })^{n-m}| vac \rangle\\ \nonumber
	=&&\frac{1}{\sqrt{n!} }\sum_{m=0}^{n} C_{n}^{m} f(x)^{m}
	f(y)^{n-m} \\ \nonumber
	&&\sqrt{m!(n-m)!}  | m \rangle_{x}| n-m \rangle_{y},\\ 
	(x,y)&&\in  \{ (H,V),(D',A'),(R',L')  \} \label{x},
\end{eqnarray}
where 
\begin{eqnarray}
	a^{\dagger } (\theta , \phi )=&&cos\frac{\theta}{2} a_{H}^{\dagger }+
	e^{i\phi  }sin\frac{\theta}{2}a_{V}^{\dagger }   \\ \nonumber 
	=&&f(H)a_{H}^{\dagger }+f(V)a_{V}^{\dagger }\\ \nonumber 
	=&&\frac{1}{\sqrt{2} }(cos\frac{\theta}{2}+e^{i(\phi -\beta )}sin\frac{\theta}{2})a_{D'}^{\dagger }+\\ \nonumber 
	&&\frac{1}{\sqrt{2} }(cos\frac{\theta}{2}-e^{i(\phi -\beta )}sin\frac{\theta}{2})a_{A'}^{\dagger }\\ \nonumber 
	=&&f(D')a_{D'}^{\dagger }+f(A')a_{A'}^{\dagger }\\ \nonumber 
	=&&\frac{1}{\sqrt{2} }(cos\frac{\theta}{2}-ie^{i(\phi -\beta )}sin\frac{\theta}{2})a_{R'}^{\dagger }+\\ \nonumber 
	&&\frac{1}{\sqrt{2} }(cos\frac{\theta}{2}+ie^{i(\phi -\beta )}sin\frac{\theta}{2})a_{L'}^{\dagger }\\ \nonumber 
	=&&f(R')a_{R'}^{\dagger }+f(L')a_{L'}^{\dagger },\\ \nonumber
\end{eqnarray}
and 
\begin{eqnarray}
	&&|f(H)| ^{2}=\frac{1+cos\theta }{2} ,|f(V)| ^{2}=\frac{1-cos\theta }{2} ,\\ \nonumber 
	&&|f(D')| ^{2}=\frac{1+sin\theta cos(\phi-\beta)}{2} ,\\ \nonumber 
	&&|f(A')| ^{2}=\frac{1-sin\theta cos(\phi-\beta)}{2}, \\ \nonumber 
	&&|f(R')| ^{2}=\frac{1+sin\theta sin(\phi-\beta)}{2} ,\\ \nonumber 
	&&|f(L')| ^{2}=\frac{1-sin\theta sin(\phi-\beta)}{2}. \\ \nonumber 
\end{eqnarray}

 According to Eq. (\ref{n}), the probability that it generates the $| m \rangle_{x}| n-m \rangle_{y}$ state is $P_{m,n-m}=C_{n}^{m}$ $| f(x) |^{2m}  | f(y) |^{2(n-m)}  $.

Then, the probability that the $n$-photon state causes the detector $x$ to respond is:
\begin{eqnarray}
	Y_{n,x}=&&\sum_{m=0}^{n} P_{m,n-m}Y_{m,n-m}^{x}\\ \nonumber 
=&&\sum_{m=0}^{n}C_{n}^{m}| f(x) |^{2m}  | f(y) |^{2(n-m)}\\\nonumber 
&&(1-(1-\eta_{d})^{m}(1-Pd))(1-\eta_{d})^{n-m}(1-Pd)\\\nonumber 
=&&(1-Pd)(| f(x) |^{2}+(1-\eta_{d})| f(y) |^{2})^{n}-\\\nonumber 
&&(1-Pd)^{2}(1-\eta_{d})^{n}\\\nonumber 
=&&(1-Pd)(1-\eta_{d}| f(y) |^{2})^{n}-(1-Pd)^{2}(1-\eta_{d})^{n},
\end{eqnarray}
and the probability that  the detector $y$  responds is:
\begin{eqnarray}
Y_{n,y}=(1-Pd)(1-\eta_{d}| f(x) |^{2})^{n}-(1-Pd)^{2}(1-\eta_{d})^{n},\quad\quad
\end{eqnarray}
where $\eta_{d}$ is the detection efficiency of the detector, and $Pd$ is the dark count rate.

The overall gain Q can be calculated as:
\begin{eqnarray}
	Q_{\xi_{B}}=&&Q_{x}+Q_{y}, \\ \nonumber 
	Q_{x}=&&\sum_{n=0}^{\infty } P(I\eta_{c},n )Y_{n,x}\\ \nonumber 
	=&&\sum_{n=0}^{\infty }e^{-I\eta_{c} }\frac{(I\eta_{c}) ^{n}}{n!}\\ \nonumber 
	&&((1-Pd)(1-\eta_{d}| f(y) |^{2})^{n}-(1-Pd)^{2}(1-\eta_{d})^{n})\\ \nonumber 
	=&&(1-Pd)e^{-I\eta_{c} \eta_{d}| f(y) |^{2}}-(1-Pd)^{2}e^{-I\eta_{c} \eta_{d}},\\ \nonumber 
	Q_{y}=&&(1-Pd)e^{-I\eta_{c} \eta_{d}| f(x) |^{2}}-(1-Pd)^{2}e^{-I\eta_{c} \eta_{d}},
\end{eqnarray}
where the subscript $\xi_{B}$ represents the measurement basis of Bob.

Then, the error rate of state $x$ will be:
\begin{eqnarray}
	E_{\xi_{B},x}=\frac{(e_{d}Q_{x}+(1-e_{d})Q_{y})}{Q_{\xi_{B}}} ,
\end{eqnarray}
where $e_{d}$ is the  optical intrinsic error rate.

Since the photons generated by the passive source are not perfect and a certain deviation is allowed, we need to integrate the over gain and the error rate within the post-selection interval as:
\begin{eqnarray}
	&&\langle  Q_{\xi_{B}} \rangle _{S_{\xi_{A}}^{y}}=\frac{1}{\langle  P \rangle _{S_{\xi_{A}}^{y}}} \iiint_{S_{\xi_{A}}^{y}} Q_{\xi_{B}}f(I,\theta,\phi)dId\theta d\phi, \quad\quad \label{Q}\\ 
	&&\langle  E_{\xi_{B}} \rangle _{S_{\xi_{A}}^{y}}=\frac{1}{\langle  P \rangle _{S_{\xi_{A}}^{y}}} \iiint_{S_{\xi_{A}}^{y}} E_{\xi_{B},x} f(I,\theta,\phi)dId\theta d\phi \label{E}, \quad\quad\\
	&&\langle P  \rangle _{S_{\xi_{A}}^{y}}=\iiint_{S_{\xi_{A}}^{y}}  f(I,\theta,\phi)dId\theta d\phi,  \label{P}
\end{eqnarray}
where $\langle P  \rangle _{S_{\xi_{A}}^{y}}$ is the normalization coefficient. $f(I,\theta,\phi)$ is the probability density function that describes the distribution of the output state on the Bloch sphere, and \cite{FP3}
\begin{eqnarray}
	f(I,\theta ,\phi )&&=f(I,\theta  )f(\phi),\\ \nonumber
	f(\phi)&&=\frac{1}{2\Delta\phi   } ,\\ \nonumber
	f(I,\theta  )&&=\frac{1}{vt\pi ^{2}\sqrt{1-\frac{I}{2vt}cos^{2}\frac{\theta}{2}  } \sqrt{1-\frac{I}{2vt}sin^{2}\frac{\theta}{2}  }} .
\end{eqnarray}

\subsection{Parameter estimation}
In practical QKD system, we need to estimate the error rate and the overall gain based on the statistical results of the detector responses. Different from the case of an infinitely long key, under the condition of a finite key length, there are usually some fluctuations in this statistical result. Regarding the measurement as a series of Bernoulli random variables, its statistical deviation can be described by Kato's inequality \cite{FP4}, and we have:
\begin{eqnarray}
	&K_{N,\epsilon}^{L}(\Lambda_{N})\overset{\epsilon} {\le}  S_{N} \overset{\epsilon} {\le} K_{N,\epsilon}^{U}(\Lambda_{N}),& \label{KUL}\\ 
	&K_{N,\epsilon}^{L}(\Lambda_{N})=\Lambda_{N}-(b+a(2\Lambda_{N}/N-1))\sqrt{N},&  \label{KL}\\ 
	&K_{N,\epsilon}^{U}(\Lambda_{N})=\Lambda_{N}+(d+c(2\Lambda_{N}/N-1))\sqrt{N},&  \label{KU}
\end{eqnarray}
where $S_{N}$ represents the expected value, $\Lambda_{N}$ represents the actual statistical value, and $\epsilon$  represents the confidence level. Eq. (\ref{KUL})  indicates that the probability that $S_{N} $ lies between $K_{N,\epsilon}^{L}(\Lambda_{N})$ and $K_{N,\epsilon}^{U}(\Lambda_{N})$ is more than $1-\epsilon$, and $\epsilon$ is usually an extremely small number. $N$ is the total number of statistical counts. $a,b,c,d$ are coefficients related to $N, \Lambda_{N}, \epsilon$, and the specific formula is given in the Appendix B.

Similarly, through the expected value, we can also estimate the upper and lower bounds of the statistical values, given by \cite{FP4}: 
\begin{eqnarray}
	&\bar{K} _{N,\epsilon}^{L}(S_{N})\overset{\epsilon} {\le}  \Lambda_{N}\overset{\epsilon} {\le} \bar{K}_{N,\epsilon}^{U}(S_{N}),&   \\ 
	&\bar{K} _{N,\epsilon}^{L}(S_{N})=(\sqrt{N}S_{N}+N(e-f))/(2e+\sqrt{N}),&  \label{K1L}\\ 
	&\bar{K} _{N,\epsilon}^{U}(S_{N})=(\sqrt{N}S_{N}+N(g-h))/(\sqrt{N}-2g).&  \label{K1U}
\end{eqnarray}
$e,f,g,h$ are coefficients related to $N, S_{N}, \epsilon$, and the specific formula is given in the Appendix B.

According to the above formulas, we can estimate the parameters in Eq. (\ref{l}). The following are the details of the estimation.

\subsubsection{ the lower bound of $M_{S_{Z}^{s}Z,1}$}
$M_{S_{Z}^{s}Z,1}$ is the single-photon response count, and it is defined as:
\begin{eqnarray}
	M_{S_{Z}^{s}Z,1}=N\langle P  \rangle _{S_{Z}^{s}} P_{Z} \langle P(I,1)  \rangle _{S_{Z}^{s}}
	\langle Y_{Z}^{1}  \rangle _{S_{Z}^{s}}^{L} \label{M}
\end{eqnarray}
where $N$ is the total number of pulses generated by the fully passive source.  $\langle P  \rangle _{S_{Z}^{s}}$ is the probability that Alice generates photons in the $Z$ basis, and can be calculated with Eq. (\ref{P}).  $P_{Z}$ is the probability that Bob selects the $Z$ basis for measurement. $\langle P(I,1)  \rangle _{S_{Z}^{s}}$ is the probability that the source generates single photons, and
\begin{eqnarray}
	\langle P(I,n)  \rangle _{S_{\xi_{A}}^{y}}=\iiint_{S_{\xi_{A}}^{y}} \frac{I^{n}e^{-I}}{n!}  f(I,\theta,\phi)dId\theta d\phi.  \label{Pn}
\end{eqnarray}
$\langle Y_{Z}^{1}  \rangle _{S_{Z}^{s}}^{L}$ is the yield of single-photon state, and 
can be estimated with the decoy state method. We can solve the following linear programming problem \cite{FP4}:
\begin{eqnarray}
&&\langle Y_{\xi _{B}}^{1}  \rangle _{S_{\xi _{A}}^{s}}^{L}=	min  \langle Y_{\xi _{B}}^{1}  \rangle _{S_{\xi _{A}}^{s}},  \label{yl}   \\ 
&&	\langle Y_{\xi _{B}}^{1}  \rangle _{S_{\xi _{A}}^{s}}^{U}=	max  \langle Y_{\xi _{B}}^{1}  \rangle _{S_{\xi _{A}}^{s}}, \label{yu}\\ \label{y}
	s.t.&&\langle \overline{Q_{\xi _{B}}}\rangle_{S_{\xi _{A}}^{j}} \ge \sum_{n=0}^{n_{cut}} \langle P(I,n)\rangle_{S_{\xi _{A}}^{j}} \langle Y_{\xi _{B}}^{n} \rangle_{S_{\xi _{A}}^{j}},\\  \nonumber
	&&\langle\overline{Q_{\xi _{B}}}\rangle_{S_{\xi _{A}}^{j}} \le \sum_{n=0}^{n_{cut}} \langle P(I,n)\rangle_{S_{\xi _{A}}^{j}} \langle Y_{\xi _{B}}^{n}  \rangle_{S_{\xi _{A}}^{j}}
	+\\  \nonumber
	&& \quad\quad\quad\quad\quad\quad\quad  1-\sum_{n=0}^{n_{cut}} \langle P(I,n)\rangle_{S_{\xi _{A}}^{j}},\\  \nonumber
	&&|   \langle Y_{\xi _{B}}^{n}  \rangle _{S_{\xi _{A}}^{i}}-\langle Y_{\xi _{B}}^{n}  \rangle _{S_{\xi _{A}}^{j}}|
	\le D(\sigma_{S_{\xi _{A}}^{i}}^{n}, \sigma_{S_{\xi _{A}}^{j}}^{n}), n=2,...,n_{cut},\\  \nonumber
	&&\langle Y_{\xi _{B}}^{n}  \rangle _{S_{\xi _{A}}^{i}}=\langle Y_{\xi _{B}}^{n}  \rangle _{S_{\xi _{A}}^{j}}, n=0,1,\\  \nonumber
	&&0\le \langle Y_{\xi _{B}}^{n}  \rangle_{S_{\xi _{A}}^{j}} \le 1 , n=0,1,...,n_{cut},\\ \nonumber 
	&&\frac{K_{N_{j},\epsilon}^{L}(M_{S_{\xi _{A}}^{j}\xi_{B}})}{N\langle P  \rangle _{S_{\xi _{A}}^{s}} P_{\xi _{B}}} \le 
	\langle\overline{Q_{\xi _{B}}}\rangle_{S_{\xi _{A}}^{j}}\le \frac{K_{N_{j},\epsilon}^{U}(M_{S_{\xi _{A}}^{j}\xi_{B}})}{N\langle P  \rangle _{S_{\xi _{A}}^{s}} P_{\xi _{B}}},\\ \nonumber 
	&& \quad\quad\quad\quad\quad\quad\quad\quad
	i,j\in \{v,d,s\},
\end{eqnarray}
where $ D(\sigma_{S_{\xi _{A}}^{i}}^{n}, \sigma_{S_{\xi _{A}}^{j}}^{n})$ is the trace distance, and is equal to $\sum |\lambda_{i}| $, where $\lambda_{i}$ are the eigenvalues of density matrix  ($\sigma_{S_{\xi _{A}}^{i}}^{n}- \sigma_{S_{\xi _{A}}^{j}}^{n}$).
\begin{eqnarray}
\sigma_{S_{\xi _{A}}^{i}}^{n}=\iiint_{S_{\xi_{A}}^{i}} |n\rangle\langle n|_{a^{\dagger}(\theta,\phi)}f(I,\theta,\phi)dId\theta d\phi.
\end{eqnarray}
$N_{j}=N\langle P  \rangle _{S_{\xi _{A}}^{j}} P_{\xi_{B}}$, and $M_{S_{\xi _{A}}^{j}\xi_{B}}=N\langle P  \rangle _{S_{\xi _{A}}^{j}}  P_{\xi_{B}}$ $ \langle Q_{\xi_{B}}\rangle_{S_{\xi _{A}}^{j}}$.
When $\xi _{A}=\xi _{B}=Z$, and $j=s$,  the solution of the linear programming Eq. (\ref{yl}) is $\langle Y_{Z}^{1}  \rangle _{S_{Z}^{s}}^{L}$. In this specific scenario, by leveraging Eq. (\ref{K1L})  in conjunction with Eq. (\ref{M}), we are able to determine the lower bound of $M_{S_{Z}^{s}Z,1}$ under the condition of a finite length as:
\begin{eqnarray}
	M_{S_{Z}^{s}Z_{B},1}^{L}= \bar{K} _{N_{j},\epsilon}^{L}(M_{S_{Z}^{s}Z_{B},1}).
\end{eqnarray}

\subsubsection{ the upper bound of $IE_{1}$}
According to Eqs. (\ref{IE1}), (\ref{u}), and (\ref{v}), $IE_{1}$ is a function that is related to $C$ and $e_{ZZ}^{1}$. We need to estimate the upper bound of 
$e_{ZZ}^{1}$ and the lower bound of $C$.
According to Eq. (\ref{C}), the lower bound of $C$ can be calculated as:
\begin{eqnarray}
	C^{L}=C^{L}_{X_{A}X_{B}}+C^{L}_{X_{A}Y_{B}}+C^{L}_{Y_{A}X_{B}}+C^{L}_{Y_{A}Y_{B}},
\end{eqnarray}
where 
\begin{eqnarray}
	C^{L}_{\xi_{A}\xi_{B}}=\left\{\begin{matrix}
		(1-2e_{\xi_{A}\xi_{B}}^{1,U})^{2},  & if \quad e_{\xi_{A}\xi_{B}}^{1,L}\ge 0.5, \\ 
		(1-2e_{\xi_{A}\xi_{B}}^{1,L})^{2},   & if \quad e_{\xi_{A}\xi_{B}}^{1,U}\le  0.5,\\
		0,  & otherwise.
	\end{matrix}\right.
\end{eqnarray}
$e_{\xi_{A}\xi_{B}}^{1,U}$ and $e_{\xi_{A}\xi_{B}}^{1,L}$ are respectively the upper bound and the lower bound of $e_{\xi_{A}\xi_{B}}^{1}$, and can be calculated as
\begin{eqnarray}
	e_{\xi_{A}\xi_{B}}^{1,U}=\frac{m_{S_{Z}^{s}Z_{B},1}^{U}}{M_{S_{Z}^{s}Z_{B},1}^{L} } ,\\ \nonumber 
	e_{\xi_{A}\xi_{B}}^{1,L}=\frac{m_{S_{Z}^{s}Z_{B},1}^{L}}{M_{S_{Z}^{s}Z_{B},1}^{U} } ,
\end{eqnarray}
where 
\begin{eqnarray}
	&&m_{S_{\xi_{A}}^{s}\xi_{B},1}^{U(L)}=\bar{K} _{N,\epsilon}^{U(L)}(m^{U(L)}), \\ \nonumber 
	&&m^{U(L)}=N\langle P  \rangle _{S_{\xi_{A}}^{s}} P_{\xi_{B}} \langle P(I,1)  \rangle _{S_{\xi_{A}}^{s}}
	\langle e_{\xi_{B}}^{1}  Y_{\xi_{B}}^{1}  \rangle _{S_{\xi_{A}}^{s}}^{U(L)},\\ \nonumber 
&&M_{S_{\xi_{A}}^{s}\xi_{B},1}^{U(L)}=\bar{K} _{N,\epsilon}^{U(L)}(M^{U(L)}) ,\\ \nonumber 
&&	M^{U(L)}=N\langle P  \rangle _{S_{\xi_{A}}^{s}} P_{\xi_{B}} \langle P(I,1)  \rangle _{S_{\xi_{A}}^{s}}
	\langle   Y_{\xi_{B}}^{1}  \rangle _{S_{\xi_{A}}^{s}}^{U(L)}.\\ \nonumber 
\end{eqnarray}

$\langle   Y_{\xi_{B}}^{1}  \rangle _{S_{\xi_{A}}^{s}}^{U(L)}$ can be obtained by solving the linear programming problem of Eqs. (\ref{yl}) and (\ref{yu}). $\langle e_{\xi_{B}}^{1}  Y_{\xi_{B}}^{1}  \rangle _{S_{\xi_{A}}^{s}}^{U(L)}$ can be estimated with the following linear programming problem \cite{FP4}:
\begin{eqnarray}
&&\langle e_{\xi_{B}}^{1}  Y_{\xi_{B}}^{1}  \rangle _{S_{\xi_{A}}^{s}}^{L}=	min   \langle e_{\xi _{B}}^{1}Y_{\xi _{B}}^{1}  \rangle _{S_{\xi _{A}}^{s}} ,\\ 
&&\langle e_{\xi_{B}}^{1}  Y_{\xi_{B}}^{1}  \rangle _{S_{\xi_{A}}^{s}}^{U}=	max   \langle e_{\xi _{B}}^{1}Y_{\xi _{B}}^{1}  \rangle _{S_{\xi _{A}}^{s}}, \\  \label{ey}
	s.t.&&\langle\overline{E_{\xi _{B}}Q_{\xi _{B}}}  \rangle_{S_{\xi _{A}}^{j}} \ge \sum_{n=0}^{n_{cut}} \langle P(I,n)\rangle_{S_{\xi _{A}}^{j}} \langle e_{\xi _{B}}^{n}Y_{\xi _{B}}^{n}  \rangle_{S_{\xi _{A}}^{j}},\\  \nonumber
	&&\langle\overline{E_{\xi _{B}}Q_{\xi _{B}} }\rangle_{S_{\xi _{A}}^{j}} \le \sum_{n=0}^{n_{cut}} \langle P(I,n)\rangle_{S_{\xi _{A}}^{j}} \langle e_{\xi _{B}}^{n}Y_{\xi _{B}}^{n}   \rangle_{S_{\xi _{A}}^{j}}
	+\\ \nonumber 
	&& \quad\quad\quad\quad\quad\quad\quad\quad 1-\sum_{n=0}^{n_{cut}} \langle P(I,n)\rangle_{S_{\xi _{A}}^{j}},\\  \nonumber
	&&|   \langle e_{\xi _{B}}^{n}Y_{\xi _{B}}^{n}  \rangle _{S_{\xi _{A}}^{i}}-\langle e_{\xi _{B}}^{n}Y_{\xi _{B}}^{n}  \rangle _{S_{\xi _{A}}^{j}}|
	\le D(\sigma_{S_{x}^{i}}^{n}, \sigma_{S_{x}^{j}}^{n}),\\  \nonumber
	&&  \quad\quad\quad\quad\quad\quad\quad\quad n=1,...,n_{cut},\\  \nonumber 
	&&\langle e_{\xi _{B}}^{0}Y_{\xi _{B}}^{0}  \rangle _{S_{\xi _{A}}^{i}}=\langle e_{\xi _{B}}^{0}Y_{\xi _{B}}^{0}  \rangle _{S_{\xi _{A}}^{j}},\\  \nonumber
	&&0\le \langle e_{\xi _{B}}^{n}Y_{\xi _{B}}^{n}  \rangle_{S_{\xi _{A}}^{j}} \le 1 , n=0,1,...,n_{cut},\\ \nonumber 
	&&\frac{K_{N_{j},\epsilon}^{L}(m_{S_{\xi _{A}}^{j}\xi_{B}})}{N\langle P  \rangle _{S_{\xi _{A}}^{s}} P_{\xi _{B}}} \le 
	\langle\overline{E_{\xi _{B}}Q_{\xi _{B}}} \rangle_{S_{\xi _{A}}^{j}}\le \frac{K_{N_{j},\epsilon}^{U}(m_{S_{\xi _{A}}^{j}\xi_{B}})}{N\langle P  \rangle _{S_{\xi _{A}}^{s}} P_{\xi _{B}}},\\ \nonumber 
	&& \quad\quad\quad\quad\quad\quad\quad\quad i,j\in \{v,d,s\},
\end{eqnarray}
where $x$ is defined in Eq. (\ref{x}), $S_{x}^{j}$ represents the post-selection interval of state $x$, and 
$m_{S_{\xi _{A}}^{j}\xi_{B}}=N\langle P  \rangle _{S_{\xi _{A}}^{j}}$ $  P_{\xi_{B}}\langle E_{\xi _{B}} Q_{\xi_{B}}\rangle_{S_{\xi _{A}}^{j}}$.

\begin{figure*}[htbp]
	\centering
	\begin{tabular}{@{}cc@{}}
		
		\includegraphics[width=0.48\linewidth]{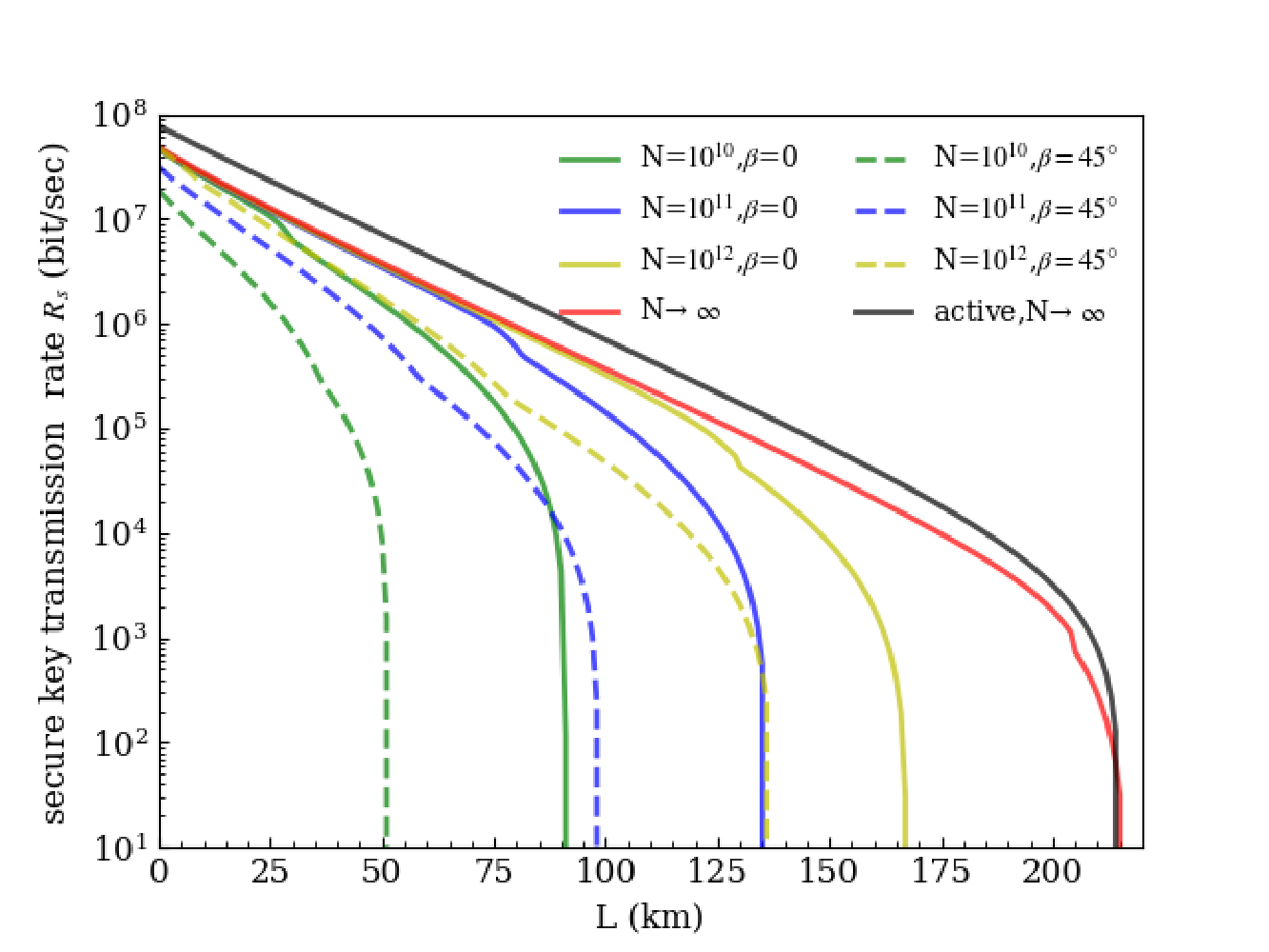} &
		\includegraphics[width=0.48\linewidth]{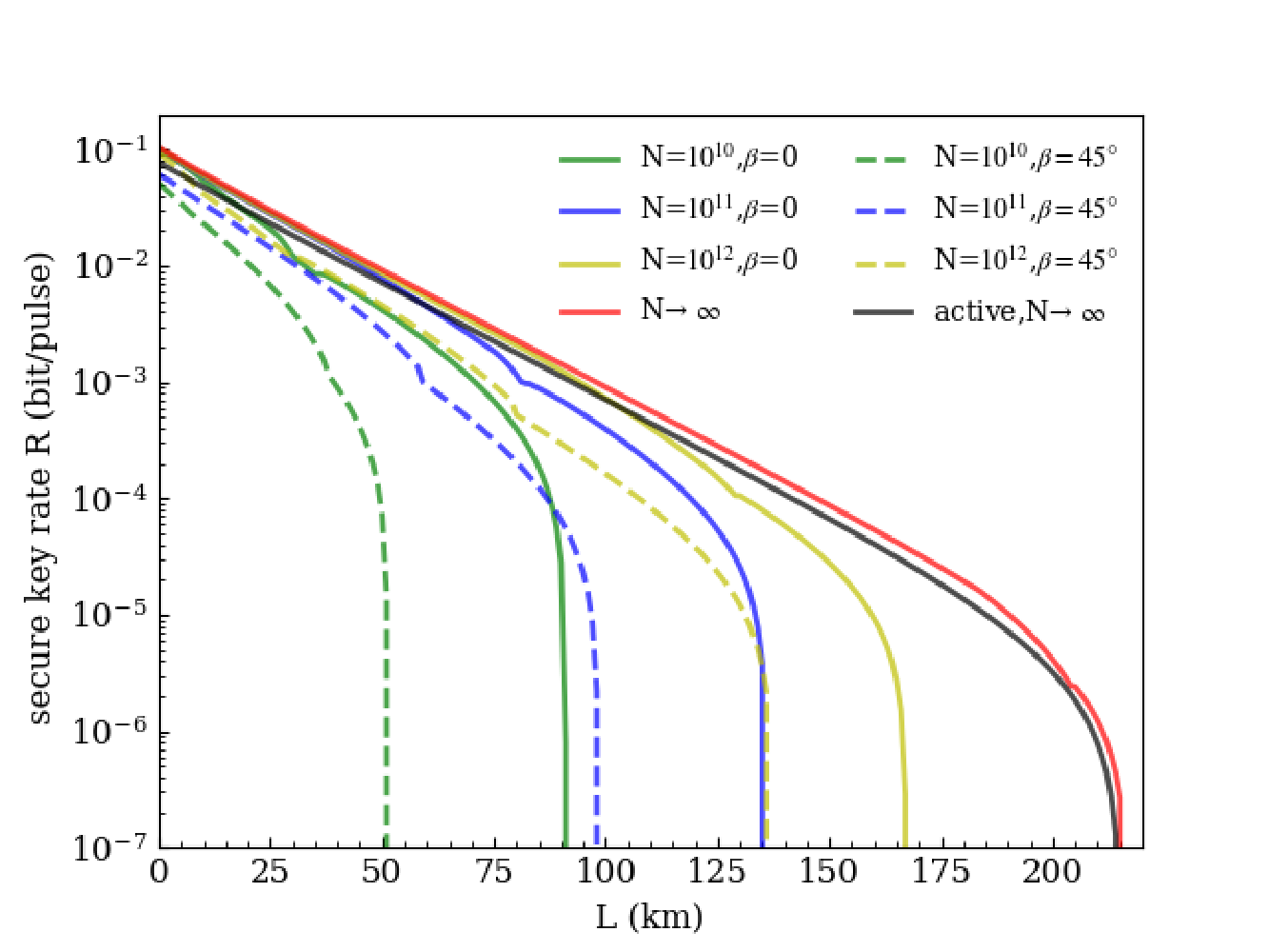} \\
		(a) Bit rate per second ($\beta=0$) & (b) Bit rate per pulse ($\beta=0$) \\[6pt]
		
	\end{tabular}
	
	\caption{ 	
		\justifying 
		The performance of our fully passive RFI QKD protocol under different numbers of pulse rounds $N$ ($N$ is the total number of pulses emitted by the fully passive source). (a) Under the optimal interval sizes and light source intensities, the maximum secure key rate that the protocol can transmit per unit time with a GHz source. (b) The key rate that can be transmitted by a single pulse in the channel,when the optimal transmission rate in Fig. (a) is achieved. 
		The black curve represents the performance of actively modulated QKD with optimal intensities and infinite length. The red curve represents the performance of the fully passive RFI QKD protocol  in an ideal case (with infinite decoy states and infinite length).
		The three solid (dashed) lines in red, yellow and blue  respectively correspond to the performance of the fully passive RFI QKD with $N = 10^{10}, 10^{11}$, and $ 10^{12}$ when the reference frame misalignment $\beta$ is 0 ($45^{\circ}$). 
	}
	\label{performance}
\end{figure*}

\subsubsection{the upper bound of $\lambda _{EC}$}
$\lambda _{EC}$ is the information entropy that needs to be consumed for error correction and $\lambda _{EC}=f_{e}M_{S_{Z}^{s}Z}^{U}h(E_{S_{Z}^{s}Z}^{U})$.
Through Eqs. (\ref{KL}) and (\ref{KU}), it is easy for us to conclude that:
\begin{eqnarray}
	M_{S_{Z}^{s}Z}^{U}&&=K_{N_{j},\epsilon}^{U}(M_{S_{Z}^{s}Z}), \\ 
	E_{S_{Z}^{s}Z}^{U}&&=\frac{m_{S_{Z}^{s}Z}^{U}}{M_{S_{Z}^{s}Z}^{L}},
\end{eqnarray}
where 
\begin{eqnarray}
	m_{S_{Z}^{s}Z}^{U}&&=K_{N_{j},\epsilon}^{U}(m_{S_{Z}^{s}Z}),\\ \nonumber 
	M_{S_{Z}^{s}Z}^{L}&&=K_{N_{j},\epsilon}^{L}(M_{S_{Z}^{s}Z}), \\ \nonumber 
M_{S_{Z}^{s}Z}&&=N\langle P\rangle _{S_{Z}^{s}}P_{Z}\langle Q_{Z}\rangle_{S_{Z}^{s}},\\ \nonumber 
	m_{S_{Z}^{s}Z}&&=N\langle P  \rangle _{S_{Z}^{s}}  P_{Z} \langle E_{Z}Q_{Z}\rangle_{S_{Z}^{s}}.
\end{eqnarray}

\section{Numerical simulation}

In this section, we analyze the performance of our protocol under finite-length conditions through numerical simulations. 
The simulation parameters we used are the same as those in  \cite{FP4}, including the detector detection efficiency $\eta_{d}=65\% $, the fiber attenuation coefficient $\alpha=0.2 \quad dB/km$, and the dark count rate $Pd=10^{-6}$.  The  optical intrinsic error rate $e_{d}$ is set as $1.5\%$. When conducting the decoy-state analysis, we set the intensity intervals of the three decoy states as $vac= [0,I_{v}] ,d= (I_{v},I_{d}] ,s = (I_{d},I_{s}] $, and $I_{v}=0.05I$, $I_{d}=0.1I$, $I_{s}=I $. 
 And the parameter $n_{cut}$ in Eqs. (\ref{y}) and (\ref{ey}) is set to be 8. We set the probability that Bob chooses the $Z$-basis for measurement as $\frac{1}{2}$, and the probabilities of choosing the $X$-basis and the $Y$-basis for measurement are both set as $\frac{1}{4}$.  In the finite-length analysis, $\epsilon_{PA}$, $\epsilon_{cor}$, and $\delta$ are set as $10^{-20}$.

Fig. \ref{performance} presents the optimal performance curves of our protocol under different scenarios and we compare our protocol  with the  actively modulated QKD. 
 The black curve represents the optimal performance of actively modulated QKD with infinite decoy states and  infinite length. The red curve represents the performance of the fully passive RFI QKD protocol in an ideal case (with infinite decoy states and infinite length). 
The three solid (dashed) lines in red, yellow and blue  respectively correspond to the performance of the fully passive RFI QKD with $N = 10^{10}, 10^{11}$, and $ 10^{12}$ when the reference frame misalignment $\beta$ is 0 ($45^{\circ}$). 
We define a  secure key transmission rate, which is expressed as $R_{s}=n\langle P\rangle _{S_{Z}^{s}}P_{Z}R$. This parameter represents the amount of key that the protocol can generate using a  source with a frequency of $n$ within a unit of time. Here, $n$ is the pulse frequency of the  source, $\langle P\rangle _{S_{Z}^{s}}P_{Z}$ is the probability that Alice and Bob can form a key, and $R$ represents the information entropy of a single pulse which is equal to $\frac{l}{N}$.   
We have considered two misalignment values of the reference frames, optimized the intervals and intensities of the protocol, and obtained the maximum secure key transmission rate of the protocol for different amounts of data in Fig. \ref{performance}.(a). 
 And the  pulse frequency of the light source is set to be 1 GHz.
 
 In the case of an infinite length and infinite decoy states, the performance of the fully passive RFI QKD can approximately reach more than 50\% of that of the actively modulated QKD.  The difference is mainly due to the fact that a part of the quantum states generated by the light source will be discarded because of post-selection. And the post-selection interval will inevitably introduce a certain error rate.
However, its maximum communication distance is slightly more than that of the active protocol, which can reach 215 km (the actively modulated protocol is 214 km).
 As the number of pulse rounds decreases, due to statistical deviations, its maximum communication distance and secure key transmission rate gradually decrease. At a distance of 80 km, the secure key transmission rates with $N=10^{12}$, $10^{11}$, $10^{10}$ when $\beta=0$  are 91.8\%, 63.2\%, and 9.9\%   of that in the case of an infinite length, respectively. And the maximum communication distance  with $N=10^{12}$, $10^{11}$, $10^{10}$ when $\beta=0$ are 
$167$, $135$, $91$ km.
Fig. \ref{performance}.(b) characterizes  the secure key rate when the optimal transmission rate in Fig. \ref{performance}.(a) is achieved. Among them, the fully passive RFI-QKD achieves a secure key rate on par with—or even marginally surpasses—that of actively modulated QKD. This advantage stems from the fact that the RFI protocol utilizes the measurement outcomes of the $X$ and $Y$ bases to bound eavesdropping, enabling tighter security constraints against Eve.

We also present the optimal secure key transmission rate curve of the protocol when the reference frame misalignment is $45^{\circ}$. Theoretically, we use the $\beta$-independent statistic $C$ to estimate eavesdropping, and the performance of the protocol should be independent of the reference frame misalignment value. However, due to the fact that the decoy state method fails to accurately estimate the single-photon error rate, and with the influence of the finite length, the actual value of $C$ exhibits a periodicity related to $\beta$. When the reference frame misalignment is $45^{\circ}$, $C$ reaches its minimum value, and at this time, the amount of information Eve obtains is the largest. 
As $N$ decreases, both the maximum communication distance and the secure key transmission rate experience significant drops. 
However, the maximum communication distance  when $N=10^{12}$, $10^{11}$, $10^{10}$ can still reach to  136,   98, 51  km, respectively.

\section{Conclusion}
In this paper, we propose a fully passive RFI QKD protocol. By passively generating the initial state, our protocol can effectively resist side-channel attacks targeting the source.  Leveraging the RFI framework,  our protocol merely  needs to align one direction of the reference frame. This not only substantially reduces the protocol's complexity but also fortifies the overall communication security.
A notable challenge associated with fully passive sources is the elevated consumption of quantum states.
 In the original fully passive QKD protocol, only two sets  of bases are required for communication. After post-selection, a significant portion of the quantum states will be discarded, thereby diminishing the protocol's efficiency.
In our protocol, we employs  the $X$ and $Y$ bases for eavesdropping estimation and the $Z$ base for key generation, which can maximize the utilization of the quantum states produced by the fully passive source. As a result, it effectively curtails potential eavesdropping by Eve, increases the secure key rate $R$, and subsequently enhances the secure key transmission rate $R_{s}$.
Under ideal circumstances, the secure key transmission rate of our protocol can reach more than 50\% of that of the ideal QKD. Under practical conditions, we have considered the finite-length effect. When the number of pulses generated by the  source reaches $10^{12}$,  when $\beta=0 $ and $\beta=45^{\circ} $, the maximum communication distances of the protocol can reach 167 km and 136 km respectively. For a GHz  source, this amount of data can be achieved in approximately several minutes.
We believe that our protocol can contribute to the development of practical QKD systems.

\section{Appendix A: the fully passive source}
\begin{figure}[!htbp]
 	\begin{center}
 		\includegraphics[width=7cm,angle=0]{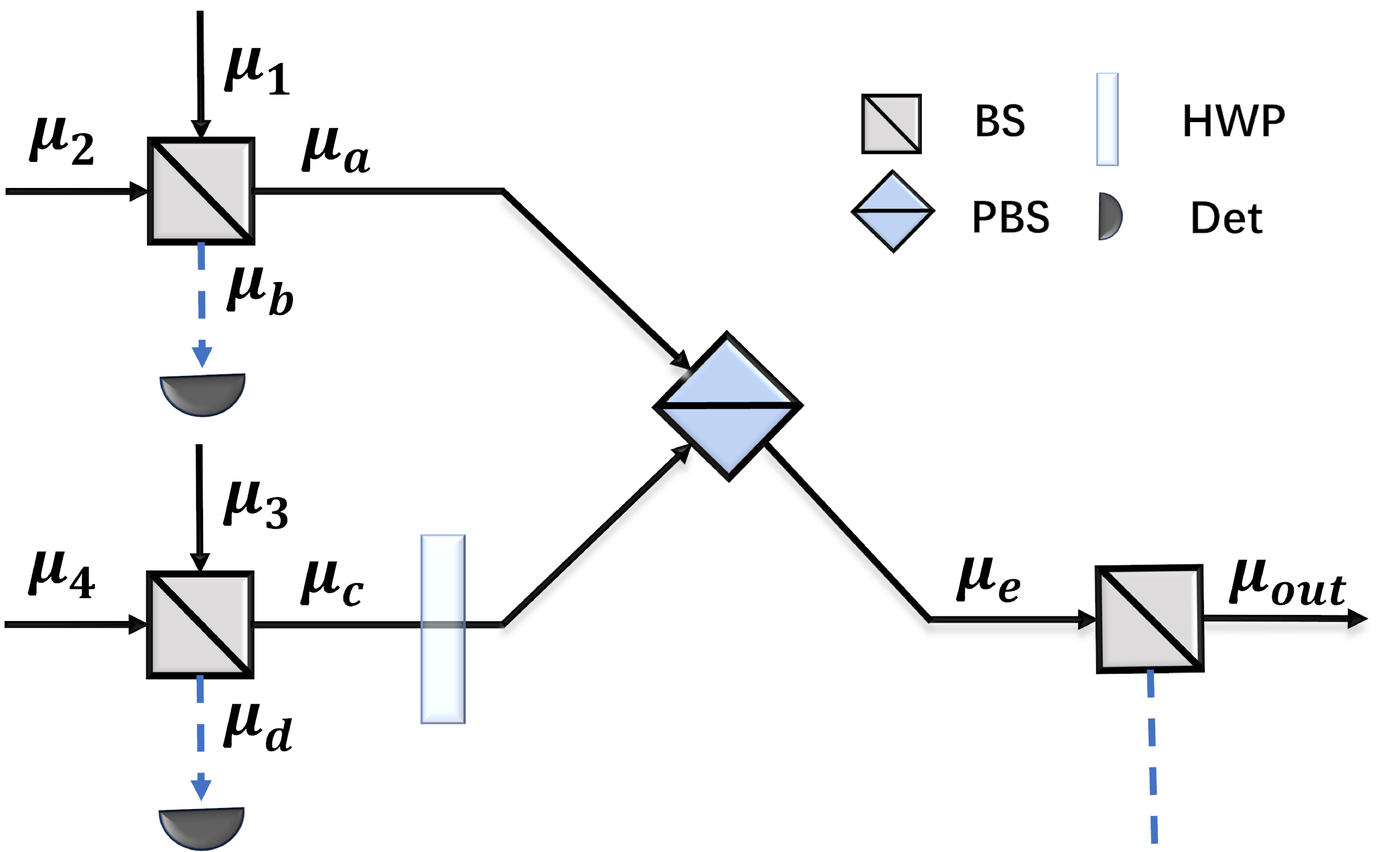}
 		\caption{\justifying
 			The structure of the fully passive source, which is similar to that in \cite{FP1}. BS: beam splitter; PBS:  polarization beam splitter; HWP: half-wave plate; Det: detector. 
 		}\label{passivesource}
 	\end{center}
 \end{figure}

In this section, we will introduce the structure of the fully passive light source we used.
In 2023, Wang et al. proposed the first fully passive source \cite{FP1}. In the same year, Zapatero et al. made several improvements \cite{FP3}. Our analysis builds upon  the researches of their respective works. We refer to \cite{FP4} and combine the coding strategy in \cite{FP1} with the analysis method in \cite{FP3}.
 By inputting four coherent lights with the same intensity and random phases, the device is capable of generating coherent lights with random quantum states and random intensities. The structural diagram of the source is shown in Fig. \ref{passivesource}. 
  
  Firstly, the four input coherent pulses can be expressed as:
\begin{eqnarray}
	| \mu_{1}   \rangle &=& | \sqrt{v}e^{i\alpha }    \rangle _{H},| \mu_{2}   \rangle = | \sqrt{v}e^{i\beta  }    \rangle _{H},\\  \nonumber
	| \mu_{3}   \rangle &=& | \sqrt{v}e^{i\gamma  }    \rangle _{H},| \mu_{4}   \rangle =| \sqrt{v}e^{i\delta   }    \rangle _{H},
\end{eqnarray}
where $v$ is the intensity, and $\alpha, \beta, \gamma,$ and $\delta $ are random phases. The subscript $H$ indicates that the photons in the pulse are initially all horizontally polarization ($| H   \rangle$). 

Then, these four coherent pulses are input into the $50:50$ beam splitter (BS) in pairs for interference, and the result is:
  \begin{eqnarray}
  	&&| \mu_{1}   \rangle| \mu_{2}   \rangle =| \sqrt{v}e^{i\alpha }    \rangle _{H} | \sqrt{v}e^{i\beta  }  \rangle _{H}\\ \nonumber
  	&&\overset{BS}{\rightarrow}| \sqrt{\frac{v}{2} }(e^{i\alpha} -e^{i\beta} )   \rangle _{H}
  	| \sqrt{\frac{v}{2} }(e^{i\alpha} +e^{i\beta} )   \rangle _{H} \\  \nonumber
  	&&=| \sqrt{2v }e^{i\frac{\alpha+\beta+\pi }{2} } sin\frac{\alpha-\beta}{2}     \rangle _{H}
  	| \sqrt{2v }e^{i\frac{\alpha+\beta}{2} } cos\frac{\alpha-\beta}{2}     \rangle _{H}
  	 \\ \nonumber
  	&&=| \mu_{a}   \rangle| \mu_{b}   \rangle, 
  	\end{eqnarray}
  	\begin{eqnarray}
  &&| \mu_{3}   \rangle| \mu_{4}   \rangle =| \sqrt{v}e^{i\gamma }    \rangle _{H} | \sqrt{v}e^{i\delta  }  \rangle _{H}\\ \nonumber
 &&\overset{BS}{\rightarrow}| \sqrt{\frac{v}{2} }(e^{i\gamma} -e^{i\delta} )   \rangle _{H}
 | \sqrt{\frac{v}{2} }(e^{i\gamma} +e^{i\delta} )   \rangle _{H} \\  \nonumber
 &&=| \sqrt{2v }e^{i\frac{\gamma+\delta+\pi }{2} } sin\frac{\gamma-\delta}{2}     \rangle _{H}
 | \sqrt{2v }e^{i\frac{\gamma+\delta}{2} } cos\frac{\gamma-\delta}{2}     \rangle _{H}
 \\ \nonumber
 &&=| \mu_{c}   \rangle| \mu_{d}   \rangle. \\ \nonumber
  \end{eqnarray}
  
Then, $| \mu_{c}   \rangle$ passes through a HWP to change the polarization state of the photons to vertical polarization ($| V   \rangle$).
$| \mu_{a}   \rangle$ and $| \mu_{c}   \rangle$  are further  input into the BS for beam combination, and we have:
\begin{eqnarray}\label{s1}
	&&| \mu_{a}   \rangle| \mu_{c}   \rangle \overset{PBS}{\rightarrow}| \mu_{e}   \rangle\\\nonumber
	=&&| \sqrt{2v }e^{i\frac{\alpha+\beta+\pi }{2} } sin\frac{\alpha-\beta}{2}     \rangle _{H}
	| \sqrt{2v }e^{i\frac{\gamma +\delta+\pi }{2} } sin\frac{\gamma -\delta}{2}     \rangle _{V}\\\nonumber
	=&&exp(-vsin^{2}\frac{\alpha-\beta}{2})exp(\sqrt{2v }e^{i\frac{\alpha+\beta+\pi }{2} } sin\frac{\alpha-\beta}{2} a_{H}^{\dagger } )\\ \nonumber
	&&exp(-vsin^{2}\frac{\gamma -\delta}{2})exp(\sqrt{2v }e^{i\frac{\gamma +\delta+\pi }{2} } sin\frac{\gamma -\delta}{2} a_{V}^{\dagger } )\\\nonumber
	=&&exp(-v(sin^{2}\frac{\alpha-\beta}{2}+sin^{2}\frac{\gamma -\delta}{2}))\\ \nonumber
	&&exp(\sqrt{2v} e^{i\frac{\alpha+\beta+\pi }{2}(sin\frac{\alpha-\beta}{2} a_{H}^{\dagger }+e^{i(\frac{\gamma +\delta}{2}-\frac{\alpha+\beta}{2})}sin\frac{\gamma -\delta}{2} a_{V}^{\dagger })})| vac   \rangle  \\\nonumber
	=&&e^{-\frac{I_{e}}{2} }e^{\sqrt{I_{e}}e^{i\psi }a^{\dagger } (\theta , \phi ) }| vac   \rangle\\ \nonumber
	=&&| \sqrt{I_{e}}e^{i\psi }  \rangle_{a^{\dagger } (\theta , \phi )},
\end{eqnarray}
where
\begin{eqnarray}
&&	I_{e}=2v(sin^{2}\frac{\alpha-\beta}{2}+sin^{2}\frac{\gamma -\delta}{2}),\\\nonumber
&&	\psi =\frac{\alpha +\beta+\pi }{2},\\\nonumber
&&	a^{\dagger } (\theta , \phi )=cos\frac{\theta }{2}a_{H}^{\dagger }+e^{i\phi}sin\frac{\theta }{2}a_{V}^{\dagger },\\\nonumber
&&	cos\frac{\theta }{2} =sin\frac{\alpha-\beta}{2}/ \sqrt{sin^{2}\frac{\alpha-\beta}{2}+sin^{2}\frac{\gamma -\delta}{2}},   \\\nonumber
&&	sin\frac{\theta }{2} =sin\frac{\gamma -\delta}{2}/ \sqrt{sin^{2}\frac{\alpha-\beta}{2}+sin^{2}\frac{\gamma -\delta}{2}},   \\\nonumber
&&	\phi =\frac{\gamma +\delta}{2}-\frac{\alpha+\beta}{2}.
\end{eqnarray}

Finally,  $| \mu_{e}   \rangle$ is attenuated by a BS whose transmission coefficient $t$ is much smaller than 1, resulting in the output of a weak coherent pulse $| \mu_{out}   \rangle$, and
\begin{eqnarray}
	| \mu_{out}   \rangle&&=e^{-\frac{I}{2} }e^{\sqrt{I}e^{i\psi }a^{\dagger } (\theta , \phi ) }| vac   \rangle\\ \nonumber
	&&=| \sqrt{I}e^{i\psi }  \rangle_{a^{\dagger } (\theta , \phi )},
\end{eqnarray}
where $I=I_{e}t$.

Since the finally output states are randomly distributed on the Bloch sphere, we still need to perform  post-selection to select the desired output states. Through heterodyne detection of $| \mu_{b}   \rangle$ and $| \mu_{d}   \rangle$, we can measure their light intensities and global phases, thereby obtaining the parameters $I, \theta,\phi$ of the output states. And the post-selection intervals are set as:
 \begin{eqnarray}
	S_{Z}^{y}=\{&&\phi \in( 0 , 2 \pi ) ,\\\nonumber
	&&\theta \in \{ ( \theta _{x} -\bigtriangleup \theta_{Z}   , \theta _{x} +\bigtriangleup \theta_{Z} ) \cap (0,\pi)\},\\\nonumber
	&&y\in \{vac,d,s\}, \theta _{x} \in \{0, \pi\}
	\},
\end{eqnarray}
	\begin{eqnarray}
	S_{X/Y}^{y}=\{&&\phi \in( \phi _{x} -\bigtriangleup \phi _{X/Y} , \phi _{x} +\bigtriangleup \phi _{X/Y} ) ,\\\nonumber
	&&\theta \in ( \frac{\pi}{2}  -\bigtriangleup \theta _{X/Y}   , \frac{\pi}{2} +\bigtriangleup \theta_{X/Y}  ) ,\\\nonumber
	&&y\in \{vac,d,s\}, \phi _{x} \in \{0, \pi\} or \{\frac{\pi}{2}, \frac{3\pi}{2} \}
	\}.
\end{eqnarray}
where the subscript of $S$ indicates the basis selection, and the superscript of $S$ indicates the decoy state interval.  $\bigtriangleup \theta_{Z}  $, $\bigtriangleup \theta_{X/Y} $ and $\bigtriangleup \phi_{X/Y}$ characterize the size of the interval. When $\theta _{x}=0 (\pi)$, the output state is denoted as  $H$ ($V$). When $\phi _{x}$ takes the values $0,\pi,\frac{\pi}{2}, \frac{3\pi}{2} $, the corresponding  output states are denoted as $D$, $A$, $R$, and  $L$.
 Among them, 
$vac= [0,I_{v}] ,d= (I_{v},I_{d}] ,s = (I_{d},I_{s}] $, and $I_{v}$, $I_{d}$, $I_{s} $ are the boundaries of three intervals.

\begin{widetext}
\section{Appendix B: the Kato's inequality}
Let $\xi_{1},..., \xi_{N}$ be a sequence of Bernoulli random variables and let $\mathcal{F}_{1}\subseteq  \mathcal{F}_{2} \subseteq ...\subseteq  \mathcal{F}_{N}$ 
be an increasing chain of $\sigma-$algebras. 
Let $\Lambda_{l}=\sum _{u=0}^{l}\xi_{u}$. The Kato's inequality \cite{kato} is shown as:

\begin{eqnarray}
	&&Pr{\Huge[}\sum _{u=1}^{N}Pr(\xi _{u}=1|\mathcal{F}_{u-1} )-\Lambda_{N}]\ge  
	[b+a(\frac{2\Lambda_{N}}{N}-1)] \sqrt{N} {\Huge]}
	\le 
	exp{\Huge[}\frac{-2(b^{2}-a^{2})}{(1+\frac{4a}{3\sqrt{N}} )^{2}} {\Huge]}. \label{K0}
\end{eqnarray}

Replacing $\xi_{l}\rightarrow 1-\xi_{l}$, and $a\rightarrow-a$ in Eq. (\ref{K0}),  we can derive:
\begin{eqnarray}
	&&Pr{\Huge[}\Lambda_{N}-\sum _{u=1}^{N}Pr(\xi _{u}=1|\mathcal{F}_{u-1} )]\ge  
	[b+a(\frac{2\Lambda_{N}}{N}-1)] \sqrt{N} {\Huge]}  
	\le 
	exp{\Huge[}\frac{-2(b^{2}-a^{2})}{(1-\frac{4a}{3\sqrt{N}} )^{2}} {\Huge]} .\label{K1}
\end{eqnarray}
Among them, $\Lambda_{N}$ represents the statistical value, and $\sum _{u=1}^{N}Pr(\xi _{u}=1|\mathcal{F}_{u-1} )$ represents the theoretical value.
These two formulas describe the upper and lower bounds of the statistical value and the theoretical value within a certain precision. Both the precision and the upper and lower bounds are related to the parameters $a$ and $b$. When the precision is fixed, we hope that the fluctuation of the upper and lower bounds is as small  as possible. That is, we need to solve: 
\begin{eqnarray}
	&&\min_{a,b} [b+a(\frac{2\Lambda_{N}}{N}-1)] \sqrt{N} \label{ku}, \\ \nonumber 
	&&s.t. exp{\Huge[}\frac{-2(b^{2}-a^{2})}{(1+\frac{4a}{3\sqrt{N}} )^{2}}{\Huge]}=\epsilon, b\ge |a|, 
\end{eqnarray}
and
\begin{eqnarray}
	&&\min_{a,b} [b+a(\frac{2\Lambda_{N}}{N}-1)] \sqrt{N} \label{kl}, \\ \nonumber 
	&&s.t. exp{\Huge[}\frac{-2(b^{2}-a^{2})}{(1-\frac{4a}{3\sqrt{N}} )^{2}}{\Huge]}=\epsilon, b\ge |a|,
\end{eqnarray}
The solution of Eq. (\ref{ku}) is \cite{kato1,FP4}
\begin{eqnarray}
	&&a=\frac{3\{9\sqrt{2} N(N-2\Lambda_{N})\sqrt{-ln\epsilon [9\Lambda_{N}(N-\Lambda_{N})-2Nln\epsilon ]
		} +16N^{3/2 }ln^{2}\epsilon -72\Lambda_{N}\sqrt{N}(N-\Lambda_{N})ln\epsilon \}
	}{4(9N-8ln\epsilon )[9\Lambda_{N}(N-\Lambda_{N})-2Nln\epsilon ]
	}, \\
	&&b=\frac{\sqrt{18Na^{2}-(16a^{2}-24\sqrt{N}a+9N)ln\epsilon  }
	}
	{3\sqrt{2} N} ,
\end{eqnarray}
and
\begin{eqnarray}
	&&\sum _{u=1}^{N}Pr(\xi _{u}=1|\mathcal{F}_{u-1} ) \overset{\epsilon } {> }K_{N,\epsilon}^{L}(\Lambda_{N}),\\ \nonumber 
	&&K_{N,\epsilon}^{L}(\Lambda_{N})=\Lambda_{N}-(b+a(2\Lambda_{N}/N-1))\sqrt{N}.
\end{eqnarray}
For the convenience of distinction, we set the solutions of Eq. (\ref{kl}) as $c=a_{opt}$  and $d=b_{opt}$.
The solution  is \cite{kato1,FP4}
\begin{eqnarray}
	&&c=\frac{3\{9\sqrt{2} N(N-2\Lambda_{N})\sqrt{-ln\epsilon [9\Lambda_{N}(N-\Lambda_{N})-2Nln\epsilon ]
		} -16N^{3/2 }ln^{2}\epsilon +72\Lambda_{N}\sqrt{N}(N-\Lambda_{N})ln\epsilon \}
	}{4(9N-8ln\epsilon )[9\Lambda_{N}(N-\Lambda_{N})-2Nln\epsilon ]
	}, \\
	&&d=\frac{\sqrt{18Nc^{2}-(16c^{2}+24\sqrt{N}c+9N)ln\epsilon  }
	}
	{3\sqrt{2} N}, 
\end{eqnarray}
and
\begin{eqnarray}
	&&\sum _{u=1}^{N}Pr(\xi _{u}=1|\mathcal{F}_{u-1} ) \overset{\epsilon } {< }K_{N,\epsilon}^{U}(\Lambda_{N}),\\ \nonumber 
	&&K_{N,\epsilon}^{U}(\Lambda_{N})=\Lambda_{N}-(d+c(2\Lambda_{N}/N-1))\sqrt{N}.
\end{eqnarray}
At this stage, the upper and lower bounds of the theoretical value $\sum _{u=1}^{N}Pr(\xi _{u}=1|\mathcal{F}_{u-1} )$ are determinable based on the statistical value $\Lambda_{N}$.

At this point, through reverse solving, we can also estimate the upper and lower bounds of the statistical value from the theoretical value. Let $\sum _{u=1}^{N}Pr(\xi _{u}=1|\mathcal{F}_{u-1} )=S_{N}$, and
\begin{eqnarray}
	&\bar{K} _{N,\epsilon}^{L}(S_{N})\overset{\epsilon} {\le}  \Lambda_{N}\overset{\epsilon} {\le} K_{N,\epsilon}^{U}(S_{N}),&   \\ 
	&\bar{K} _{N,\epsilon}^{L}(S_{N})=(\sqrt{N}S_{N}+N(e-f))/(2e+\sqrt{N}),&  \label{KA1L}\\ 
	&\bar{K} _{N,\epsilon}^{U}(S_{N})=(\sqrt{N}S_{N}+N(g-h))/(\sqrt{N}-2g),&  \label{KA1U}
\end{eqnarray}
where $e=a_{opt}$ and $f=b_{opt}$ are the solution of:
\begin{eqnarray}
	&&\max_{a,b} \frac{\sqrt{N}S_{N}+N(a-b)}{2a+\sqrt{N}} ,
	\label{kl1} \\ \nonumber 
	&&s.t. exp{\Huge[}\frac{-2(b^{2}-a^{2})}{(1+\frac{4a}{3\sqrt{N}} )^{2}}{\Huge]}=\epsilon, b\ge |a|,
\end{eqnarray}
and \cite{kato1,FP4}
\begin{eqnarray}
	&&e=\frac{3\sqrt{N}\{9 (N-2S_{N})\sqrt{Nln\epsilon [Nln\epsilon -18S_{N}(N-S_{N})]}-4Nln^{2}\epsilon 
		-9ln\epsilon (8S_{N}^{2}-8NS_{N}+3N^{2})\}
	}{4\{4Nln^{2}\epsilon +36ln\epsilon (2S_{N}^{2}-2NS_{N}+N^{2})+81NS_{N}(N-S_{N})\}
	}, \\
	&&f=\frac{1}{3} \sqrt{9e^{2}-\frac{(4e+3\sqrt{N})^{2}ln\epsilon  
		}{2N
		} 
	} .
\end{eqnarray}
$g=a_{opt}$ and $h=b_{opt}$ are the solution of:
\begin{eqnarray}
	&&\min_{a,b} \frac{\sqrt{N}S_{N}-N(a-b)}{\sqrt{N}-2a} ,
	\label{kl1} \\ \nonumber 
	&&s.t. exp{\Huge[}\frac{-2(b^{2}-a^{2})}{(1-\frac{4a}{3\sqrt{N}} )^{2}}{\Huge]}=\epsilon, b\ge |a|,
\end{eqnarray}
and \cite{kato1,FP4}
\begin{eqnarray}
	&&g=\frac{3\sqrt{N}\{9 (N-2S_{N})\sqrt{Nln\epsilon [Nln\epsilon -18S_{N}(N-S_{N})]}+4Nln^{2}\epsilon 
		+9ln\epsilon (8S_{N}^{2}-8NS_{N}+3N^{2})\}
	}{4\{4Nln^{2}\epsilon +36ln\epsilon (2S_{N}^{2}-2NS_{N}+N^{2})+81NS_{N}(N-S_{N})\}
	}, \\
	&&h=\frac{
		\sqrt{18Ng^{2}- (16g^{2}-24\sqrt{N}g+9N)ln\epsilon 
	}}{3\sqrt{2}N 
	} .
\end{eqnarray}
\end{widetext}

\section*{Acknowledgement}
 This work is supported by the National Natural Science Foundation of China under Grants  No. 12175106 and  No. 92365110.
 
\nocite{*}

\quad

\end{document}